\documentclass[sigconf]{acmart}

\usepackage{booktabs}
\usepackage{adjustbox}

\AtBeginDocument{%
  \providecommand\BibTeX{{%
    \normalfont B\kern-0.5em{\scshape i\kern-0.25em b}\kern-0.8em\TeX}}}

\begin{document}

\title{Machine learning applications for electricity market agent-based models: A systematic literature review}

\author{Alexander J. M. Kell}
\email{a.kell@imperial.ac.uk}
\affiliation{%
  \institution{Imperial College London}
  \city{London}
  \state{Greater London}
  \country{UK}
  \postcode{43017-6221}
}

\author{A. Stephen McGough}
\affiliation{%
  \institution{Newcastle University}
  \city{Newcastle}
  \country{UK}
}

\author{Matthew Forshaw}
\affiliation{%
  \institution{Newcastle University}
  \country{UK}}
\email{matthew.forshaw@newcastle.ac.uk}

\renewcommand{\shortauthors}{Kell et al.}

\begin{abstract}
The electricity market has a vital role to play in the decarbonisation of the energy system. However, the electricity market is made up of many different variables and data inputs. These variables and data inputs behave in sometimes unpredictable ways which can not be predicted a-priori. It has therefore been suggested that agent-based simulations are used to better understand the dynamics of the electricity market. Agent-based models provide the opportunity to integrate machine learning and artificial intelligence to add intelligence, make better forecasts and control the power market in better and more efficient ways. In this systematic literature review, we review 55 papers published between 2016 and 2021 which focus on machine learning applied to agent-based electricity market models. We find that research clusters around popular topics, such as bidding strategies. However, there exists a long-tail of different research applications that could benefit from the high intensity research from the more investigated applications.
\end{abstract}

\ccsdesc[500]{Computing methodologies~Modeling and simulation}
\ccsdesc[500]{Computing methodologies~Machine learning}
\ccsdesc[500]{Computing methodologies~Artificial intelligence}

\keywords{machine learning, multi-agent system, literature review, electricity, markets, forecasting}

\maketitle

\section{Introduction}
\label{sec:ML-abms}

It is now almost universally acknowledged that we are having a devastating impact on our environment \cite{May2002}. To help mitigate these impacts, a move towards a low-carbon energy supply will be required. A clear understanding of the mechanisms and their outcomes in the electricity generation markets is becoming increasingly important as the first step towards mitigating the possible devastating impacts of climate change. Due to the falling prices of low-carbon electricity generation technologies, such as wind and solar power \cite{IRENA2018}, electricity is increasingly being used to decarbonise energy demands such as heating, automotive and in the industrial sector.

With the advent of renewable energy supply technologies, novel methodologies are required to manage an electricity system with increased fluctuation due to the nature of many renewable intermittent generation processes. This evolution in the electricity system raises important questions, such as how to balance supply and demand, how to bid in an uncertain market and how to forecast prices. Electricity market models are an important tool used by policy makers, researchers and industry to investigate questions such as these. 

Traditional methods for analysing electricity systems, such as centralised optimisation models have certain limitations such as the requirement for perfect foresight, homogeneity of actors within the electricity system and a cost-optimal system \cite{Peng2021, Pfenninger2014, Ringler2016}. Agent-based models (ABMs) have been proposed as a novel methodology to mitigate these limitations. Specifically, agent-based models can model imperfect foresight, heterogeneity of actors with differing objectives and individual level actions can be simulated.

ABMs are increasingly incoporating artificial intelligence (AI) and machine learning (ML) to provide answers to complex questions such as electricity frequency control, demand response and bidding strategies. As this is a growing field of research, this paper attempts to provide a review of AI and ML applied to ABMs for electricity markets. The goal of this paper is to analyze the current literature on how AI is being used with ABMs in the area of electricity markets. This paper makes three separate contributions to the field: 
\begin{enumerate}
\item A classification of problem domains used by ABMs with ML and AI
\item An overview of the relevant literature which uses these techniques 
\item New avenues of research to use ABMs with AI and ML.
\end{enumerate}

This paper is structured as follows: Section \ref{sec:background} introduces the concept of ABMs with AI and ML applied to electricity markets in more detail. Section \ref{sec:analysis} introduces the methodology undertaken to review the literature. Section \ref{sec:review}, the presented literature is discussed, placed in context and compared. Section \ref{sec:future} explores a possible future research direction.

\section{Background}
\label{sec:background}

\subsection{Simulation of complex systems}

Many real-world complex systems feature heterogenous, interacting and adaptive units as well as exhibiting emergent properties \cite{Ringler2016}. The analysis, forecasting and control of these systems can prove challenging due to the many different features. In addition, the behaviour of these systems can be non-linear and feature emergent properties that can be difficult to predict. However, it remains important to perform analyses and better understand these systems to achieve the intended goals \cite{Kell2019}.

Simulation and modelling can be used to perform ``what-if'' analyses ``in-silico''. That is, through the replication of salient features of the real-world system into computer code, a model of this system can be designed. Once the model has been created, multiple different analyses can be undertaken to investigate the system, which allows practitioners to build knowledge of the complex systems which they are studying. 

Analyses can be carried out where real-world experimentation would be prohibitive. This could be due to the requirement of high costs or because the risk of experimentation would be excessively high. In addition, many experimentations can be undertaken in parallel allowing users to find the best parameter set or action to reach a particular objective.

These models can be used to investigate the effect of policies, shocks such as sudden price rises or other scenarios on the system. In many cases, these investigations would not have been possible without a simulation approach. Another use of these simulations is to generate theories and for verification. However, these models always have limitations, which should be improved on. This could be by improving implementation, calibration, verification, validation, interpretation and visualisation of results \cite{Ringler2016}. These are key challenges for any model or computer simulation.

\subsection{Agent-based models and simulation of electricity systems}

Agent-based modelling is a framework to design a particular type of a computer simulation. These models contain a number of agents with certain decision rules and relationships with other agents. The behaviour of the individual agents aggregates and leads to emergent system behaviour. This emergent system behaviour can be difficult to  define a-priori, and is why these simulations can be useful.

Electricity systems are increasingly being modelled with an agent-based modelling framework \cite{Kell2020b,Ringler2012}. This is likely a consequence of the transition from a homogenous central actor to the heterogenous nature of multiple actors within a decentralised electricity system. These heterogenous actors may have different objectives, risk tolerances and budgets, among other factors which may have a significant impact on the final electricity system. Agent-based modelling is well suited to a system such as the one described here. 

Agent-based models offer a methodology that can be used in international or national electricity systems, local energy markets or in microgrids. In this review, we present work which have used agent-based models for these three different scopes, underlying the flexibility of this approach.

\subsection{Machine learning and artificial intelligence}

Machine learning (ML) is the study of computer algorithms that improve automatically with the use of data. A computer program is said to learn from experience E with respect to some class of tasks T and performance measure P, if its performance at tasks in T, as measured by P, improves with experience E \cite{mitchell1997machine}.

Machine learning methods can be split into three different categories: (1) supervised learning, (2) unsupervised learning and (3) reinforcement learning. Each of these methods can be used in the following cases:

\begin{enumerate}
	\item \textbf{Supervised learning} is used where the data has labels, such as predicting the energy used by a sub-station where we have measurements of the known, real values.
	\item \textbf{Unsupervised learning} is where there are no labels associated with the data. The model must, therefore, infer useful patterns from the data without knowledge of meaning or significance. For example, a user may have data on two species of animals and must distinguish between the two without having data on which animal belongs to which species a-priori.
	\item \textbf{Reinforcement learning} is the case where an agent is placed within an environment and must determine the best action to take to maximise some reward. For example, an agent bidding into a market needs to know how much it should bid to maximise long-term profit.
\end{enumerate}

In addition to the aforementioned machine learning methods, there exists an additional paradigm: optimisation methods. These methods explore a mathematical or software function to find a minimum or maximum value of an objective. These can be used to minimise an expected error, minimise total cost or maximise total return from a system, for example.

In the following subsections we explore an overview of the different machine learning techniques that are applicable to the problems addressed in this review:

\subsubsection{Supervised Learning}

Supervised learning uses training data, which contains both the inputs and desired outputs, to predict the output associated with new inputs. A functioning supervised learning model, will therefore be able to correctly predict the output for inputs that were not part of the original training data. Supervised learning can be used for both regression and classification. Regression is where a continuous output is returned, whereas classification is where a discrete value is returned.

A select number of widely used supervised learning algorithms are support vector machines \cite{Cortes1995}, linear regression \cite{seal1967studies}, logistic regression \cite{walker1967estimation}, decision trees \cite{quinlan1987simplifying} and neural networks \cite{Hill1994}.

\subsubsection{Unsupervised Learning}

As previously discussed, unsupervised learning infers patterns from unlabelled data. Unsupervised learning can therefore exhibit self-organisation that can capture patterns in data that were previously unknown.

Some of the most common algorithms include hierarchical clustering \cite{murtagh2012algorithms}, \textit{k}-means clustering \cite{Hartigan1979} and self-organizing maps \cite{Kohonen1990}.




Unsupervised learning is not the preferred option when labelled data is available. Therefore it is used less in simulations and agent-based models due to the high availability of data. However, in the real-world unsupervised learning can be a powerful tool, where labelled data is lacking.

\subsubsection{Reinforcement Learning}

Reinforcement learning (RL) is an algorithm type which allows intelligent agents to take actions in an environment in order to maximise a cumulative reward. Reinforcement learning does not require labelled input/output data. The basis of RL is to find a balance between exploration and exploitation.

A large amount of research has been dedicated in recent times to improving the performance of reinforcement learning algorithms \cite{Arulkumaran2017, Hunt2016a}. Various different methodologies have been tried and tested, from the use of neural networks in deep reinforcement learning to updating a lookup table.

%
%
%
%

RL is well suited to simulation environments. This is due to the high availability of observation data, the ability to craft rewards seen by the agents and the high number of simulation iterations that can be run to train the agents. They are also powerful at dealing with uncertainty and can perform actions for a wide variety of problems. It is for these reasons that RL is used a lot in simulations, and particularly agent-based models.

\section{Analysis}
\label{sec:analysis}

Artificial intelligence (AI) and machine learning (ML) have been integrated with agent-based models to model the electricity sector with increasing frequency over the last years. This is due to the ability of AI to optimise agent behaviour, system parameters and add functionality to agent-based models (ABMs). This study, therefore, reviewed recent papers regarding applications of AI and ML in this space. We categorised these papers into four different criteria: market type, application, algorithm type and algorithm used. To do this, we used different search terms on Scopus and reviewed all 55 articles in the field over the past five years. It was found that the majority of papers used reinforcement learning applied to bidding strategies. However, a range of applications were investigated through a wide variety of means. This included price forecasting, demand forecasting, microgrid management and risk management. We highlight the major findings from this study in the following sections.


In this section, we review the literature that investigates how artificial intelligence and machine learning can be integrated into agent-based models for the electricity sector. To select the related articles to review, we conduct a systematic literature review of relevant research in the field. We limit our search to literature published in the five most recent years (2016-2021). As a result of this, we provide a comprehensive status of the applications of ML and AI in agent-based models for the electricity sector. For this purpose we used the Elsevier Scopus database. To find the articles, we used the following set of search terms to select our articles:

\begin{enumerate}
	\item Machine Learning, Artificial Intelligence, Deep Learning, Neural Networks, Decision Tree, Support Vector Machine, Clustering, Bayesian Networks, Reinforcement Learning, Genetic Algorithm, Online Learning, Linear regression.
	\item Agent-based modelling.
	\item Electricity.
\end{enumerate}

We searched using each of the keywords in each of the bullet points. For instance, the first keyword search was: Machine Learning, Agent-Based Modelling and Electricity. The second was: Artificial Intelligence, Agent-based modelling and Electricity. We selected these search terms to focus this review on agent-based models applied to the electricity sector and machine learning, which is the focus of this review.

These search terms resulted in 149 research articles. However, not all of these were related to our research focus. For instance, a number of electric vehicle, buildings and biological papers were returned. After reading the titles, followed by the abstracts, these 149 papers were reduced to 55 papers which were specifically related to agent-based modelling, electricity, artificial intelligence and machine learning.

\begin{figure}
	\centering
	\includegraphics[width=0.8\linewidth]{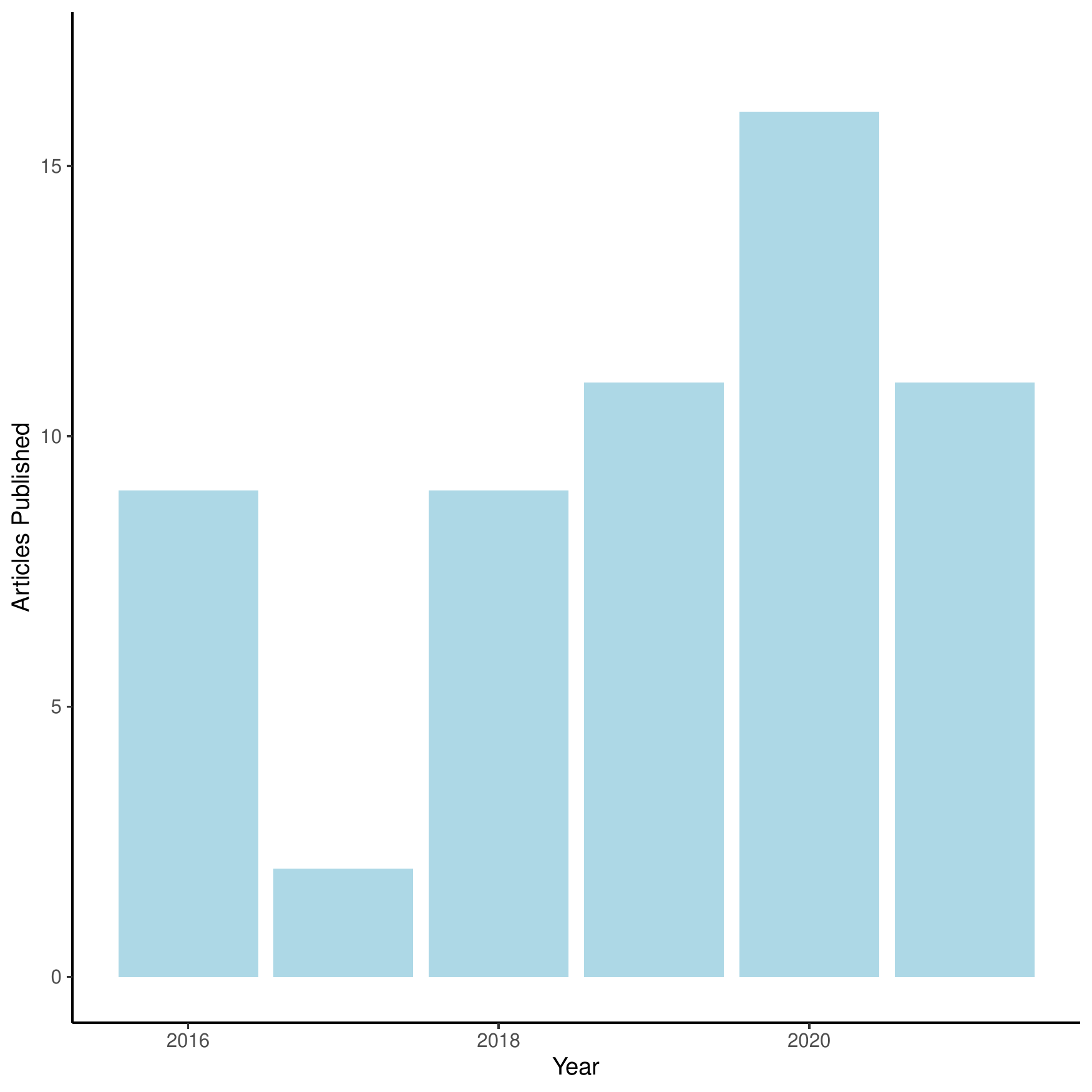}
	\caption{Number of articles published per year which work on AI in electricity focused ABMs.}
	\label{fig:articles-published}
\end{figure}

Figure \ref{fig:articles-published} shows the amount of articles published each year between 2016 and the present date. Whilst the number of articles published in this field has increased per year since 2018 to 2020, the number of papers published in 2017 was lower when compared to the other years, with a large number published in 2016. We reviewed these 55 papers systematically in the following sections.

\subsection{Market Type}

In this literature review, we make three different market type distinctions: international/national energy market, local energy market and a microgrid. The international/national energy market typically considers a country, multiple countries or the world. A local energy market is a smaller region than the international/national energy market, for instance, a city or region. Whereas a microgrid serves a discrete geographic footprint, such as a university campus, business centre or neighbourhood. Whilst there is some cross-over between a local energy market and microgrid, a microgrid can be disconnected from the traditional grid and operate autonomously. 

Tables \ref{table:RL}, \ref{table:supervised-learning} and \ref{table:unsupervised-learning}, \ref{table:optimisation} and \ref{table:game-theory} categorise  the papers to their respective market types. The papers have been displayed in chronological order and categorise the market type, machine learning (ML) type used, the application in which it was used and the algorithm used. These different criteria are explored in the following subsections.

\subsection{Machine Learning Types}

Within this work, we have covered five different types of artificial intelligence paradigms. These are: supervised learning, unsupervised learning, reinforcement learning, optimisation and game theory.

\begin{figure}
	\centering
	\includegraphics[width=0.85\linewidth]{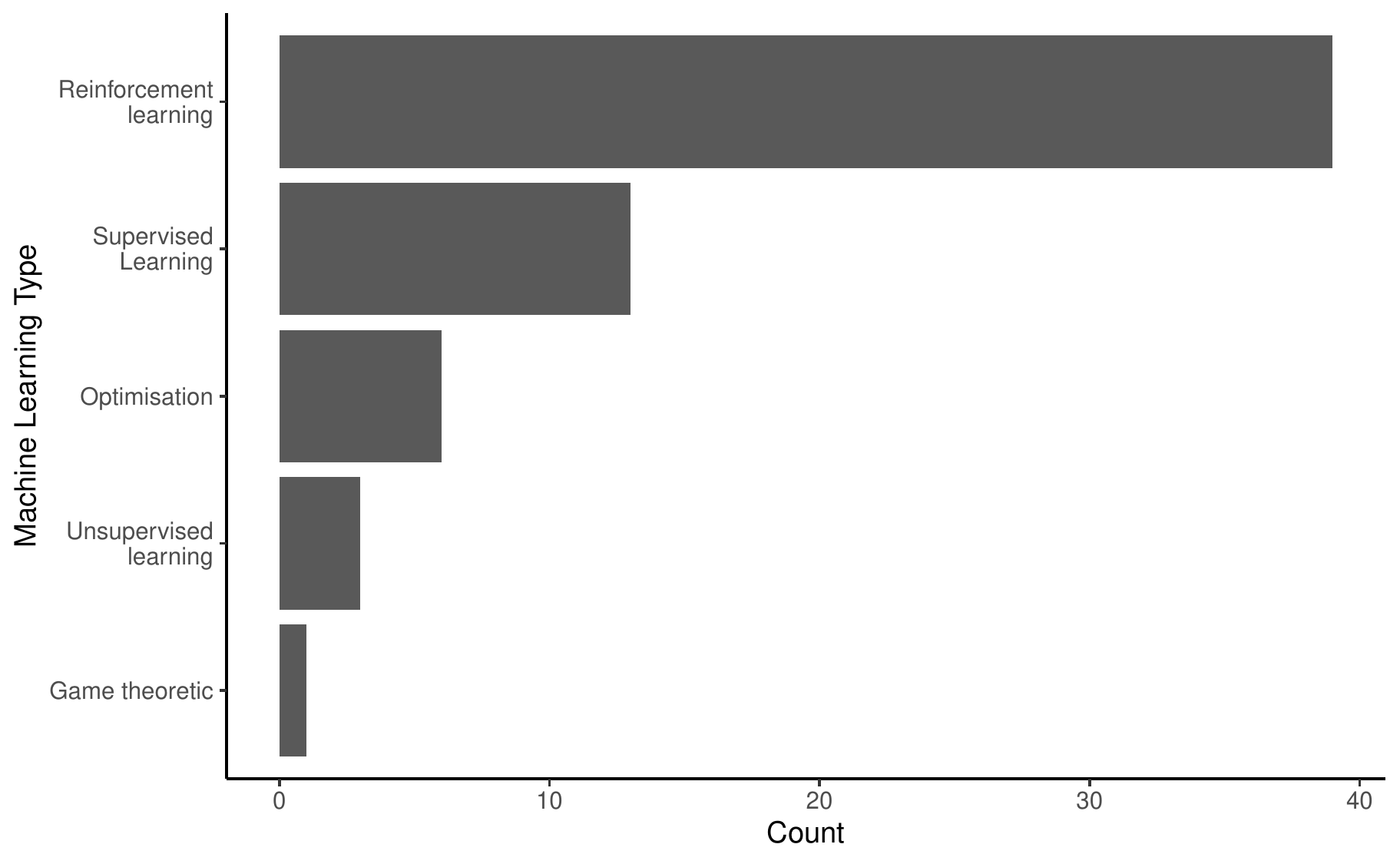}
	\caption{Number of machine learning categories surveyed in papers.}
	\label{fig:ml_type}
\end{figure}

Each of these techniques have been utilised in the papers surveyed. However, a particular prevalence has been placed on reinforcement learning within the research community. As shown by Figure \ref{fig:ml_type}, 67\% of papers used a reinforcement learning algorithm. This greatly outweighs the other machine learning types. The second most used machine learning type was supervised learning, used by eight papers. The fact that reinforcement learning has been used so extensively within the agent-based modelling community for electricity highlights the usefulness of this technique within this field, as previously discussed. 

\begin{figure*}
	\centering
	\includegraphics[width=0.85\linewidth]{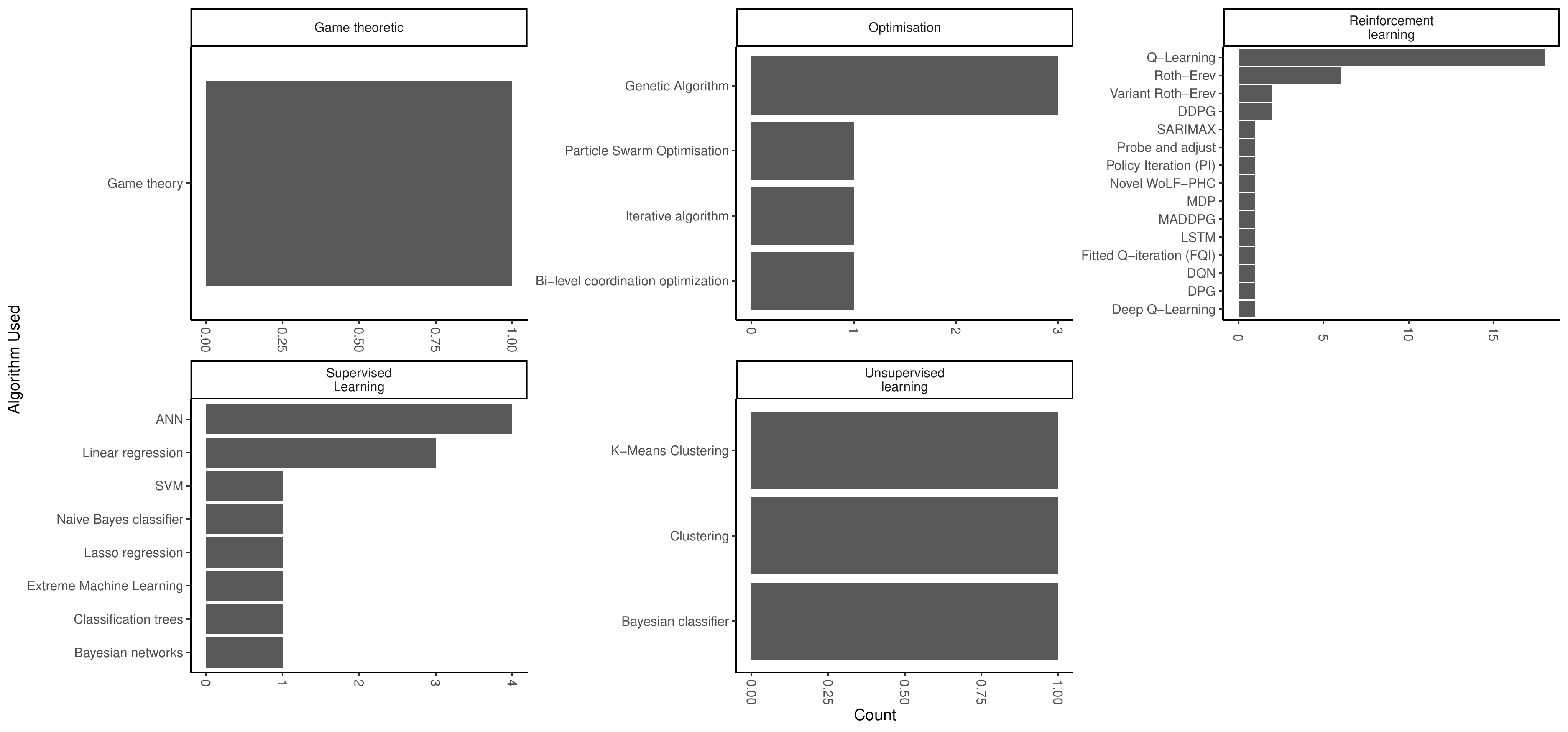}
	\caption{Machine learning algorithms used per machine learning category.}
	\label{fig:ml_type_algorithm}
\end{figure*}

Within each of the different machine learning categories there exist many algorithms. The algorithms used in the papers surveyed are presented in Figure \ref{fig:ml_type_algorithm}. Within reinforcement learning the most commonly used algorithms are Q-Learning \cite{Gay2007}, Roth-Erev and Variant Roth-Erev \cite{RothAE1995}.

Within supervised learning, the most commonly used algorithms are artificial neural networks (ANN) \cite{Hill1994}, linear regression \cite{seal1967studies} and support vector machines (SVM) \cite{Cortes1995}. Fewer algorithms were used for both unsupervised learning and optimisation. For unsupervised learning, only the following algorithms were used: Bayesian classifier \cite{Tschiatschek2014}, K-Means Clustering \cite{Hartigan1979} and Naive Bayes classifier \cite{Tschiatschek2014}. For optimisation the following algorithms were trialled: Bi-level coordination optimisation \cite{dempe2002foundations}, Genetic Algorithm. \cite{mitchel}, Iterative algorithm \cite{kelley1999iterative} and Particle Swarm Optimisation \cite{kennedy1995particle}. For the game theory method, a game theoretic algorithm was used \cite{myerson1997game}.

\subsection{Applications}

Within this work, we classified each paper by the problem domain which they are trying to solve, or the application. The applications are: agent behaviour, bidding strategies, bilateral trading, demand forecasting, demand response, electricity grid control, expansion planning, forecasting carbon emissions, load scheduling, market investigation, microgrid management, peer to peer trading, price forecasting, risk management, scheduling of flexibility, secure demand side management and tariff design. Table \cite{table:application_defs} 

\begin{table*}[]
\caption{Application definitions.}
\footnotesize
\begin{tabular}{@{}ll@{}}
\toprule
Application                   & Definition                                                                \\ \midrule
Agent behaviour               & Use of AI to control decisions made by agents                             \\
Bidding strategies            & Bid within an electricity market                                          \\
Bilateral trading             & Inform trading decisions based on two trading entities                    \\
Demand forecasting            & Forecasting of future electricity (or other) demand                       \\
Demand response               & Reduction or increase of electricity demand to fit supply                 \\
Electricity grid control      & Control of electricity flow around an electricity grid                    \\
Expansion planning            & Decisions made upon which generators should be invested in                \\
Forecasting carbon emissions  & Forecasting of carbon emissions using AI                                  \\
Load scheduling               & Logistical planning of physical power flows by an operator                \\
Market investigation          & Analysis of the wider power market                                        \\
Microgrid management          & Management of electricity flows within a microgrid                        \\
Peer to peer trading          & Strategies for trading between two or more entities within a power market \\
Price forecasting             & Forecasting of electricity (or other) prices                              \\
Risk management               & Management and reduction of risk                                          \\
Scheduling of flexibility     & Flexibility management of power system to ensure that supply meets demand \\
Secure demand side management & A resilient method to control demand to fit supply                        \\
Tariff design                 & Design of tariffs to charge or pay within a power market                  \\ \bottomrule
\end{tabular}
\label{table:application_defs}
\end{table*}

\begin{figure}
	\centering
	\includegraphics[width=0.95\linewidth]{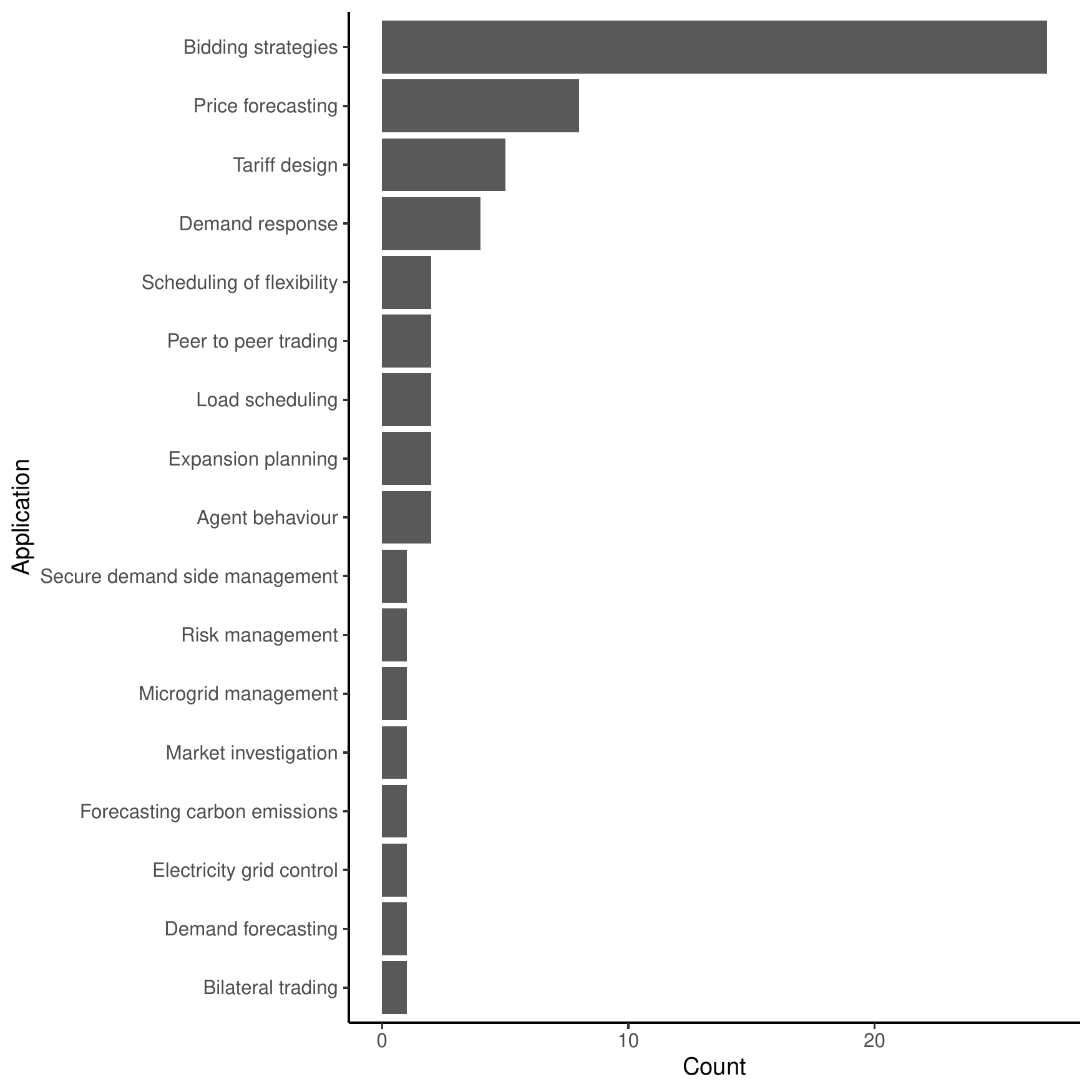}
	\caption{Application number per paper.}
	\label{fig:application}
\end{figure}

Figure \ref{fig:application} displays the number of applications used by each paper. The most utilised application was bidding strategies, with price forecasting and tariff design following behind. However, the bidding strategies application was investigated 49\% times, with price forecasting investigated only 8 times. This demonstrates a considerable research effort in this area.

\begin{figure*}
	\centering
	\includegraphics[width=0.95\linewidth]{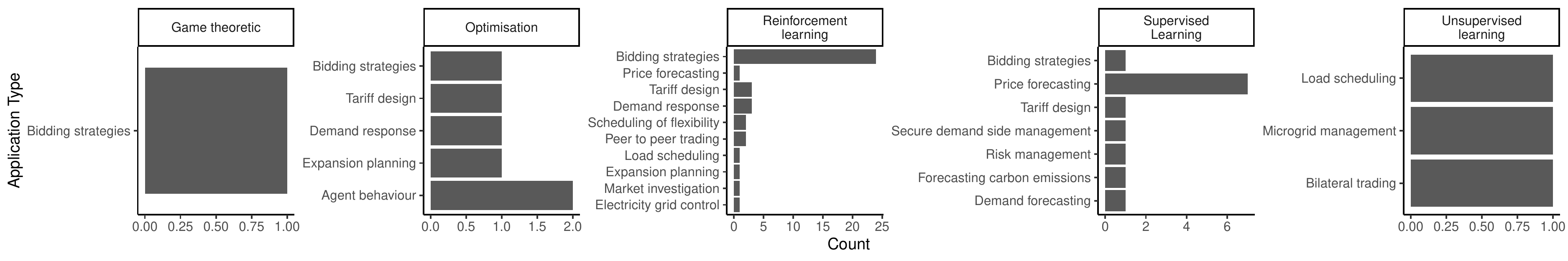}
	\caption{Applications per application type.}
	\label{fig:application-ml_type}
\end{figure*}

Figure \ref{fig:application-ml_type} displays the number of applications per machine learning type area. We can see that bidding strategies are highly used within the reinforcement learning type. The reinforcement learning algorithm, however, is shown to be highly versatile, with different applications investigated, from demand response, flexibility scheduling to expansion planning. This is due to the ability for reinforcement learning to learn different policies based upon solely the reward and observations within an environment.

Within supervised learning, a large amount of research effort has been put into price forecasting. This is likely due to the strong ability of supervised learning techniques at making predictions. However, outside of classification and making predictions, supervised learning is not so versatile, when compared to reinforcement learning.

Optimisation is used for five different applications. This is because optimisation requires a problem domain to be maximised or minimised. This may not be the case for all applications.

Unsupervised learning and the game theoretic machine learning types are used less than the other machine learning types. This is because unsupervised learning is preferential when there is no labelled data. However, with labelled data, supervised learning can yield more accurate results. Within simulations it is often the case that data is available, and so supervised learning is used in preference to unsupervised learning. The game theoretic approach has been used in a single application in the papers surveyed: bidding strategies. The application of game theory is possible for the problem of bidding strategies, however, the assumptions of a Nash equilibrium and perfect information may not always exist in an electricity market.

\subsection{Market Type}

Table \ref{table:market_type} displays the frequency of market types for the papers reviewed. Whilst Table \ref{table:market_type-application} shows the number of different applications per market type.

\begin{enumerate}
	\item 68.9\% of the papers reviewed relate to an international/national electricity market. The availability of data and the relative importance of the subject of whole system transitions in current affairs may explain why such research effort has been dedicated to this. 
	\item 32.8\% of the papers for the international/national focus on bidding strategies, as is shown by Table \ref{table:market_type-application}. This is due to the ability of reinforcement learning to make strategic decisions without full knowledge of the system. This is true within electricity markets, where bidding strategies must be formulated without the knowledge of the behaviour of other actors. In addition, the ability to model strategic bidding is of significance importance for international/national energy models due to the appearance of oligopolies in national energy markets. However, whilst many studies have explored on strategic bidding and a few have focused on oligopolistic bidding strategies, a study on the impact on the wider market does not exist.
	\item Price forecasting is the second largest category which is explored, and is investigated in 13.1\% of papers. This is due to the ability for supervised learning to make predictions in time-series data. It is also of importance for agents to make realistic predictions of future prices. However, there exists a gap in the literature on the long-term effects of the different accuracies of these forecasts.
	\item 18\% of papers focus on the local energy market. A significant reduction when compared to the international/national electricity market. This is because of the limited availability of publicly available data for these local energy markets. ABMs require a high amount of data to inform the behaviour of the agents and environment, and so data collection for local energy markets can be expensive and difficult to obtain. Microgrids are explored in 13.1\% of papers.  Similarly to local energy markets this is because of a smaller amount of publicly available data. It is also possible that researchers place an increasing focus on international/national electricity markets due to the availability of these models and perceived impact from a large system.
\end{enumerate}

\begin{table}[]
\begin{tabular}{@{}ll@{}}
\toprule
Market Type                          & Percentage (\%) \\ \midrule
International/National & 68.9            \\
Local energy market                  & 18.0            \\
Microgrid                            & 13.1            \\ \bottomrule
\end{tabular}
\caption{Frequency of market type for papers reviewed.}
\label{table:market_type}
\end{table}

\begin{table*}[]
\begin{tabular}{@{}lll@{}}
\toprule
Market Type                          & Application                   & Percentage (\%) \\ \midrule
Microgrid                            & Bidding strategies            & 4.9             \\
Microgrid                            & Scheduling of flexibility     & 3.3             \\
Microgrid                            & Electricity grid control      & 1.6             \\
Microgrid                            & Load scheduling               & 1.6             \\
Microgrid                            & Microgrid management          & 1.6             \\
Local energy market                  & Bidding strategies            & 6.6             \\
Local energy market                  & Tariff design                 & 4.9             \\
Local energy market                  & Demand response               & 3.3             \\
Local energy market                  & Forecasting carbon emissions  & 1.6             \\
Local energy market                  & Peer to peer trading          & 1.6             \\
International/National & Bidding strategies            & 32.8            \\
International/National & Price forecasting             & 13.1            \\
International/National & Agent behaviour               & 3.3             \\
International/National & Expansion planning            & 3.3             \\
International/National & Tariff design                 & 3.3             \\
International/National & Bilateral trading             & 1.6             \\
International/National & Demand forecasting            & 1.6             \\
International/National & Demand response               & 1.6             \\
International/National & Load scheduling               & 1.6             \\
International/National & Market investigation          & 1.6             \\
International/National & Peer to peer trading          & 1.6             \\
International/National & Risk management               & 1.6             \\
International/National & Secure demand side management & 1.6             \\ \bottomrule
\end{tabular}
\caption{Frequency of market type and application for papers reviewed.}
\label{table:market_type-application}
\end{table*}

\subsection{Applications}
\begin{enumerate}
	\item  A significant proportion of papers have focused on bidding strategies, with 44.3\% of papers investigating this. Bidding strategies is top for each of the market types, indicating the versatility and applicability of this technique in electricity markets.
	\item The application of control, for instance grid control, load scheduling and demand response has seen a significant amount of promising research. Particular in the application of decentralised control \cite{Najafi, Shafie-Khah, Causevic}. With the advent of distributed data collection and internet of things, it has become possible to achieve various welfare objectives through a distributed control algorithm. That is, to optimise for individuals or distributed generators from the perspective of those respective individuals and generators. Studies have shown efficiencies close to centralised algorithms. However, centralised algorithms may reduce the agency of individuals and therefore must overcome an additional barrier in uptake if the individuals do not perceive that the algorithm is working in their own best interest.
	\item Various applications have been explored within the research community. In total, 17 different applications were explored. This demonstrates the versatility of ML applied to ABMs. This, however, highlights a significant gap in the literature as the majority of applications have only been explored by one or two papers. For instance, the ability to optimise the electricity system parameters in question has not been explored to the same level of detail as forecasting or trading behaviour. 
	\item Whilst there are many studies which look at optimising certain aspects of an electricity market, there is no study which looks at optimising the whole electricity system as a whole.
\end{enumerate}

\subsection{Algorithm Type}

Table \ref{table:algorithm} displays the frequency of each of the algorithms used. Q-Learning is the most used algorithm, with 29\% of papers using this algorithm. Second is the Roth-Erev algorithm, which is used by 9.7\% of papers. Whilst this shows the versatility of these algorithms, further research could be placed into the use of deep reinforcement learning (DRL) to improve results. DRL uses deep neural networks to act as function approximators. DDPG, DQN and DPG are examples of DRL algorithms. 

The majority of the algorithms have only been used in a single paper, and so, there remains a significant gap in the literature to apply these algorithms to the different applications to investigate with results can be improved through other algorithms.

\begin{table}[]
\begin{tabular}{@{}ll@{}}
\toprule
Algorithm Used                     & Percentage (\%) \\ \midrule
Q-Learning                         & 29.0            \\
Roth-Erev                          & 9.7             \\
ANN                                & 6.5             \\
Linear regression                  & 4.8             \\
Genetic Algorithm                  & 4.8             \\
DDPG                               & 3.2             \\
Variant Roth-Erev                  & 3.2             \\
DPG                                & 1.6             \\
DQN                                & 1.6             \\
SVM                                & 1.6             \\
SARIMAX                            & 1.6             \\
Bayesian networks                  & 1.6             \\
Bi-level coordination optimization & 1.6             \\
Probe and adjust                   & 1.6             \\
Policy Iteration (PI)              & 1.6             \\
Particle Swarm Optimisation        & 1.6             \\
Novel WoLF-PHC                     & 1.6             \\
Naive Bayes classifier             & 1.6             \\
MDP                                & 1.6             \\
MADDPG                             & 1.6             \\
Classification trees               & 1.6             \\
Lasso regression                   & 1.6             \\
LSTM                               & 1.6             \\
Bayesian classifier                & 1.6             \\
Iterative algorithm                & 1.6             \\
Clustering                         & 1.6             \\
Game theory                        & 1.6             \\
Fitted Q-iteration (FQI)           & 1.6             \\
Extreme Machine Learning           & 1.6             \\
Deep Q-Learning                    & 1.6             \\
K-Means Clustering                 & 1.6             \\ \bottomrule
\end{tabular}
\label{table:algorithm}
\caption{Frequency of algorithms for papers reviewed.}
\end{table}

\section{Literature review}
\label{sec:review}

\subsection{Reinforcement Learning}

In this section we review the papers that utilised reinforcement learning for the applications shown in Figure \ref{fig:application-ml_type}. Firstly, we cover the papers which consider the bidding strategies problem.

Liu \textit{et al.}\cite{Liu2020} establish non-cooperative and cooperative game models between thermal power companies. They show that, compared with other RL algorithms, the MADDPG algorithm \cite{Lowe} is more accurate, with an increase in revenue of 5.2\%.  They show that thermal companies are more inclined to use physical retention methods to make profits in the medium and long-term power market. 

Liang \textit{et al.}\cite{Liang} use non multi-agent algorithm, called the DDPG algorithm \cite{Hunt2016a} to model the bidding strategies of GenCos. They show that the DDPG algorithm is able to converge to a Nash equilibrium even with imperfect information. They find that they are able to reflect collusion through adjusting the GenCos' patience parameter. 

Purushothaman \textit{et al.}\cite{Purushothaman} model the learning capabilities of power generators through the use of the Roth-Erev RL algorithm. They find that the agents are able to exhibit market power through this approach. Kiran \textit{et al.}\cite{Kiran} use a variant of the Roth-Erev algorithm to investigate the ability for a generator to bid strategically within a market. They find that agents have the ability to bid strategically, and increase their net earnings. However, the impact on the wider market and how to limit this ability was not explored in these papers.

Sousa \textit{et al.}\cite{Sousa} use an ABM to model the Iberian electricity market, with a focus on hydropower plants. Differently to the previously mentioned papers, they use a state-table based approach, the Q-Learning algorithm to bid into the day-ahead market. Poplavskaya \textit{et al.}\cite{Poplavskaya} model the balancing services market, and investigate the effect of different market structures on price. They find that in an oligopoly, prices can deviate from the competitive benchmark by a factor of 4-5. They conclude that changing market type would not solve this issue.

Nunna \textit{et al.}\cite{Nunna} also use a Q-learning algorithm to develop bidding strategies for energy storage systems, however their algorithm focuses on a simulated-annealing-based approach. They observe that they can effectively reinforce the balance between supply and demand in the microgrids using a mixture of local and global energy storage systems.

Differently to the previously mentioned papers, Lin \textit{et al.}\cite{Lin} investigate herding behaviours of electricity retailers on the monthly electricity market. Herding behaviours are where individuals act collectively as part of a group. They model herding behaviours mathematically based on the relationship network of electricity retailers which are imbedded in an ABM. They use the Roth-Erev RL to model these behaviours. They find that the herding behaviours might bring positive or negative effects to electricity retailers depending on the differing bidding strategies.

Wang \textit{et al.}\cite{Wang} investigate the bidding behaviour of all players in the electricity market. They propose a hybrid simulation model and integrate RL to bid. They find that with the hybrid simulation model, the dynamics of the entire market remain stable, the market clearing prices converge, and the market share is relatively uniform.

Ye \textit{et al.}\cite{Ye} propose a novel multi-agent deep RL algorithm, where they combine the DPG \cite{silver2014deterministic} algorithm with LSTM \cite{Hochreiter1997} for multi-agent intelligence. Their algorithm achieves a significantly higher profit than other RL methods for a GenCo.

Pinto \textit{et al.}\cite{Pinto1} investigate the ability for collaborative RL models to optimise energy contract negotiation strategies. The results show that the collaborative learning process enables players' to correctly identify which negotiation strategy to apply in each moment, context and opponent.

Machado \textit{et al.}\cite{MacHado} take a different approach, and investigate how the energy price is affected when a government intervention is observed through the increase in number of public companies participating in auctions. They find that they are able to model the impact of public companies on the overall electricity market, rather than trying to improve the bidding strategy of a single company.

Feng \textit{et al.}\cite{Feng} follow a similar, governmental policy approach to Machado \textit{et al.} and explore the effect of a transition from a monopoly to a competitive electricity market. They simulate a day-ahead market and use RL to bid strategically. They find that they can characterise the risk characteristics and decision-making objectives of market participants.

Calabria \textit{et al.}\cite{Calabria} simulate the Brazilian power system, using 3 years worth of data which covers 98\% of the total hydro capacity of Brazil. They use RL to simulate the behaviour of virtual reservoirs in this market. They find that through the management of these virtual reservoirs, they can save water, while maintaining current efficiency and security levels.

Gaivoronskaia \textit{et al.}\cite{Gaivoronskaia} present a modification of the classical Roth-Erev algorithm to represent agents' learning in an ABM of the Russian wholesale electricity market, in a day-ahead market. Staudt \textit{et al.}\cite{Staudt} investigate the effect of the number of energy suppliers needed for a competitive market. They find that several suppliers are required to ensure a welfare optimal pricing. Where welfare is defined as a benefit to the community.

Esmaeili Aliabadi \textit{et al.}\cite{Esmaeili} study the effect of risk aversion on GenCos' bidding behaviour in an oligopolistic electricity market. They find that the change in the risk aversion level of even one GenCo can significantly impact on all GenCo bids and profits. 

Rashedi \textit{et al.}\cite{Rashedi} use Q-Learning to learn optimal bidding strategies of suppliers in a day-ahead market. They compare the competitive behaviour of players in both the multi-agent and single-agent case. Chrysanthopoulos \textit{et al.}\cite{Chrysanthopoulos} take a similar approach, by modelling the day-ahead market as a stochastic game. They use an ABM to model GenCos to maximise their profit using RL. 

Skiba \textit{et al.}\cite{Skiba} model the bidding behaviour on the day-ahead and control reserve power markets using an RL algorithm in an ABM. They investigate the resulting market prices. Tang \textit{et al.}\cite{Tang} investigate the bidding strategies of generators under three pricing mechanisms. They find they are able to achieve market equilibrium results through their novel RL approach.

Bakhshandeh \textit{et al.}\cite{Bakhshandeh} assess the ability for GenCos to withhold capacity to increase the market price using RL. The results demonstrate the emergence of capacity withholding by GenCos, which have an effect on the market price. 

Xu \textit{et al.}\cite{Xu} simulate a proactive residential demand response in a day-ahead market using RL. This is in contrast to the general electricity markets modelled in the previously mentioned papers. They use residential data in China, and test a case with 30,000 households. They find that a proactive residential demand response may yield significant benefits for both the supply and demand side.

Deng \textit{et al.}\cite{Deng} use a DDPG algorithm to also model residential demand response.  They find that the  goal of peak load-cutting and valley filling can be achieved with this method.

Shafie-Khah \textit{et al.}\cite{Shafie-Khah} develop a novel decentralised demand response model. In their model, each agent submits bids according to the consumption urgency and a set of parameters by the RL algorithm, Q-Learning. Their studies show that their decentralised model drops the electricity cost dramatically, which was nearly as optimal as a centralised approach. Najafi \textit{et al.}\cite{Najafi} also propose a decentralised demand response model. They use Q-Learning and consider small scale GenCos. Similarly to Shafie-Khah \textit{et al.}, they show the effectiveness of this technique.

Tomin \textit{et al.}\cite{Tomin} propose an RL method to interact with the electricity grid for active grid management, as opposed to bidding strategies. They demonstrate the effectiveness of this approach on a test 77-node scheme and a real 17-node network diagram of the Akademgorodok microdistrict (Irkutsk). 

Huang \textit{et al.}\cite{Huang} propose a generation investment planning model for GenCos. They use a genetic algorithm and Q-learning to improve their optimisation ability, and show that the model is effective and could provide support for plant expansion planning. Manabe \textit{et al.}\cite{Manabe} also consider the generation expansion planning problem. They find through their simulation that their agents can generate higher profit using RL.

Foruzan \textit{et al.}\cite{Foruzan} use an ABM to study distributed energy management in a microgrid. They use an RL algorithm to allow generation resources, distributed storages and customers to develop optimal strategies for energy management and load scheduling. They are able to reach a Nash equilibrium, where all agents benefit through this approach.

Viehmann \textit{et al.}\cite{Viehmann} analyse the different markets: Uniform Pricing (UP) and Discriminatory Pricing. UP is where all agents pay the maximum accepted price, and DP pay their accepted bid. Through the use of RL, they find that UPs lead to higher prices in all analysed market settings.

Pinto \textit{et al.}\cite{Pinto} introduce a learning model to enable players to identify the expected prices of bilateral agreements as opposed to the multilateral markets surveyed by the previously mentioned papers. For this they use a Q-Learning algorithm, and they use real data from the Iberian electricity market.

Mbuwir \textit{et al.}\cite{Mbuwir} explore two model-free RL techniques: policy iteration (PI)  \cite{lagoudakis2003least} and fitted Q-iteration (FQI) \cite{riedmiller2005neural} for scheduling the operation of flexibility providers - battery and heat pump in a residential microgrid. Their simulation results show that PI outperforms FQI with a 7.2\% increase in photovoltaic self-consumption in the multi-agent setting and a 3.7\% increase in the single-agent setting. 

Naseri \textit{et al.}\cite{Naseri} propose an autonomous trading agent which aims to maximise profit by developing tariffs. They find  that designing tariffs with usage-based charges and fixed periodic charges can help the agent segment the retail market resulting in lower peak demand and capacity charges.

Bose \textit{et al.}\cite{Bose} simulate a local energy market as a multi-agent simulation of 100 households. Through the use of the Roth-Erev reinforcement learning algorithm to control trading, and demand response of electricity. They are able to achieve self-sufficiency of up to 30\% with trading, and 41.4\% with trading and demand response. Kim \textit{et al.} \cite{Kim} consider a prosumer that consumes and produces electric energy with an energy storage system. They use the DQN reinforcement learning algorithm to maximise profit in peer-to-peer energy trading. Liu \textit{et al.}\cite{Liu2021} also investigate the peer-to-peer trading problem in a local energy market. They find that the community modelled are able to increase profitability through the use of DQN RL.

\begin{table*}[]
\begin{tabular}{@{}llp{3cm}p{3cm}p{3cm}@{}}
\toprule
Year & First Author              & Market Type                          & Application                        & Algorithm Used                                 \\ \midrule
2021 & Bose S.    \cite{Bose}          & Local energy market                  & Peer to peer trading               & Roth-Erev                                      \\
2021 & Naseri N.     \cite{Naseri}       & Local energy market                  & Tariff design                      & SARIMAX, MDP                                    \\
2021 & Tang C.   \cite{Tang}           & International/National & Tariff design                      & Novel WoLF-PHC                                 \\
2021 & Liu D.    \cite{Liu2021}           & International/National & Bidding strategies, Peer-to-Peer                 & MADDPG                                         \\
2021 & Deng C. \cite{Deng}             & International/National & Demand response                    & DDPG                                           \\
2021 & Viehmann J. \cite{Viehmann}         & International/National & Market investigation               & Q-Learning                                     \\
2020 & Tomin N.   \cite{Tomin}          & Microgrid                            & Electricity grid control           & Q-Learning                                     \\
2020 & Liang Y. \cite{Liang}            & International/National & Bidding strategies                 & DDPG                                           \\
2020 & Kim J.-G.     \cite{Kim}       & International/National & Peer to peer trading               & Deep Q-Learning                                \\
2020 & Shafie-Khah M.   \cite{Shafie-Khah}    & Local energy market                  & Demand response                    & Q-Learning                                     \\
2020 & Sousa J.C. \cite{Sousa}          & International/National & Bidding strategies                 & Q-Learning                                     \\
2020 & Poplavskaya K.   \cite{Poplavskaya}    & International/National & Bidding strategies                 & Q-Learning                                     \\
2020 & Liu Y.    \cite{Liu2020}           & Local energy market                  & Bidding strategies                 & DQN                                            \\
2020 & Nunna H.S.V.S.K.  \cite{Nunna}   & Microgrid                            & Bidding strategies                 & Q-Learning                                     \\
2020 & Purushothaman K. \cite{Purushothaman}    & International/National & Bidding strategies                 & Roth-Erev                                      \\
2020 & Mbuwir B.V.  \cite{Mbuwir}        & Microgrid                            & Scheduling of flexibility          & Policy Iteration (PI), Fitted Q-iteration (FQI) \\
2020 & Kiran P.     \cite{Kiran}        & International/National & Bidding strategies                 & Variant Roth-Erev                              \\
2019 & Lin F.   \cite{Lin}            & International/National & Bidding strategies                 & Roth-Erev                                      \\
2019 & Wang J.    \cite{Wang}          & International/National & Bidding strategies                 & Roth-Erev                                      \\
2019 & Machado M.R. \cite{MacHado}        & International/National & Bidding strategies                 & Q-Learning                                     \\
2019 & Ye Y.    \cite{Ye}            & International/National & Bidding strategies                 & DPG,LSTM                                       \\
2019 & Pinto T.    \cite{Pinto}         & International/National & Price forecasting                  & Q-Learning                                     \\
2019 & Pinto T.  \cite{Pinto1}           & International/National & Bidding strategies                 & Q-Learning                                     \\
2018 & Feng H.    \cite{Feng}          & Local energy market                  & Bidding strategies                 & Variant Roth-Erev                              \\
2018 & Foruzan E.     \cite{Foruzan}      & Microgrid                            & Load scheduling                    & Q-Learning                                     \\
2018 & Calabria F.A.   \cite{Calabria}     & International/National & Bidding strategies                 & Q-Learning                                     \\
2018 & Gaivoronskaia E.A. \cite{Gaivoronskaia}   & International/National & Bidding strategies                 & Roth-Erev                                      \\
2018 & Staudt P.       \cite{Staudt}     & Microgrid                            & Bidding strategies                 & Probe and adjust                               \\
2018 & Najafi S.    \cite{Najafi}        & Local energy market                  & Demand response, Bidding strategies & Q-Learning                                     \\
2017 & Esmaeili Aliabadi D. \cite{Esmaeili} & International/National & Bidding strategies                 & Q-Learning                                     \\
2016 & Huang X.    \cite{Huang}         & International/National & Expansion planning                 & Q-Learning                                     \\
2016 & Rashedi N.   \cite{Rashedi}        & International/National & Bidding strategies                 & Q-Learning                                     \\
2016 & Skiba L.    \cite{Skiba}         & International/National & Bidding strategies                 & Roth-Erev                                      \\
2016 & Bakhshandeh H.   \cite{Bakhshandeh}    & International/National & Bidding strategies                 & Q-Learning                                     \\
2016 & Chrysanthopoulos N. \cite{Chrysanthopoulos} & Local energy market                  & Bidding strategies                 & Q-Learning                                     \\ \bottomrule
\end{tabular}
\caption{Articles relating to reinforcement learning algorithm type.}
\label{table:RL}
\end{table*}

\subsection{Supervised Learning}

In this section we review the papers that used a supervised learning approach with their agent-based models, which focus on electricity. First we consider papers which focus on price forecasting.

Fraunholz \textit{et al.}\cite{Fraunholz} use ANNs to forecast electricity price endogenously within the long-term energy model, PowerACE. They find that the ANN method outperforms the linear regression method and that this endogenous method has a significant impact on simulated electricity prices. This is of importance since these are major results for electricity market models. Pinto \textit{et al.}\cite{Pinto3} uses SVMs and ANNs for price forecasting using real data from MIBEL, the Iberian market operator. They show an ability to return promising results in a fast execution time.

Goncalves \textit{et al.}\cite{Goncalves} use multiple different methods to understand the main drivers of electricity prices. These include lasso and standard regression \cite{Tibshirani1996}, and causal analysis such as Bayesian networks and classification trees. Their results are coherent and show the impact of different generators on final electricity price.

Opalinski \textit{et al.}\cite{Opalinski} propose a hybrid prediction model, where the best results from a possible large set of different short-term load forecasting models are automatically selected based on their past performance by a multi-agent system. They show an increase in prediction accuracy with their approach. This is in contrast to the previous papers which trial different approaches and manually select algorithms based on the results.

Bouziane \textit{et al.}\cite{Bouziane} forecast carbon emissions using a hybrid ANN and ABM approach from different energy sources from a city. They forecast energy production using agents and calculate the benefits of using renewable energy as an alternative way of meeting electricity demand. They find they are able to reduce emissions by 3\% per day using this approach.

Pinto \textit{et al.}\cite{Pinto4} propose an approach to addressing the adaptation of players' behaviour according to participation risk. To do this, they combine the two most commonly used approaches of forecasting: internal data analysis and sectorial data analysis. They show that their proposed approach is able to outperform most market participation strategies and reach a higher accumulated profit by adapting players' actions according to the participation risk.

Maqbool \textit{et al.}\cite{Maqbool} investigate the impact of feed-in tariffs and the installed capacity of wind power on electricity consumption from renewables. They use linear regression to understand the outputs from an agent-based model to achieve this. They find that the effect of increasing installed capacity of wind power is more significant than the effect of feed-in tariffs.

El Bourakadi \textit{et al.}\cite{Bourakadi} propose the use of an Extreme Machine Learning  (EML) \cite{huang2006extreme} algorithm to make decisions about selling/purchasing electricity from the main grid and charging and discharging batteries from an ABM. The EML algorithm predicts wind and photovoltaic power output from weather data and then makes a classification decision on whether to buy or sell power. 

Babar \textit{et al.}\cite{Babar} propose a secure demand-side management engine to preserve the efficient utilisation of energy based on a set of priorities. A model is developed to control intrusions into the smart grid using a naive Bayes classifier. Simulation is used to test the efficiency of the system, and the result reveal that the engine is less vulnerable to intrusion.

\begin{table*}[]
\begin{tabular}{@{}llp{3cm}lp{3cm}@{}}
\toprule
Year & First Author         & Market Type                          & Application                  & Algorithm Used                                                            \\ \midrule
2021 & Fraunholz C. \cite{Fraunholz}   & International/National & Price forecasting            & ANN                                                                       \\
2021 & Bouziane S.E.  \cite{Bouziane} & Local energy market                  & Forecasting carbon emissions & ANN                                                                       \\
2020 & Babar M.  \cite{Babar}   & International/National & Secure demand side management & Naive Bayes classifier \\
2019 & Maqbool A.S.   \cite{Maqbool} & International/National & Tariff design                & Linear regression                                                         \\
2019 & Goncalves C. \cite{Goncalves}   & International/National & Price forecasting            & Linear regression, Lasso regression, Bayesian networks, Classification trees \\
2019 & Pinto T.   \cite{Pinto4}     & International/National & Risk management              & ANN                                                                       \\
2019 & El Bourakadi D. \cite{Bourakadi} & Microgrid                            & Bidding strategies           & Extreme Machine Learning                                                  \\
2016 & Opalinski A.  \cite{Opalinski}  & International/National & Demand forecasting           & Linear regression                                                         \\
2016 & Pinto T.  \cite{Pinto3}      & International/National & Price forecasting            & ANN, SVM                                                                   \\ \bottomrule
\end{tabular}
\caption{Articles relating to supervised learning algorithm type.}
\label{table:supervised-learning}
\end{table*}

\subsection{Unsupervised learning}

In this section we discuss the papers which use an unsupervised learning approach with agent-based models.

Imran \textit{et al.}\cite{Imran} develop a novel strategy for bilateral negotiations. They enable each GenCo to estimate the reservation price of its opponent using Bayesian learning. They show that the agents which use Bayesian learning gain an advantage over non-learning agents. 

Čaušević \textit{et al.}\cite{Causevic} propose a novel clustering algorithm to cluster agents into virtual cluster members for the application of load scheduling. They show that large-scale centralised energy systems can operate in a decentralised fashion when only local information is available.

Gomes \textit{et al.}\cite{Gomes} propose a management system for the operation of a microgrid by an electricity market agent. They use K-means clustering for scenario reduction and a stochastic mixed-integer linear programming problem to manage the system.

\begin{table*}[]
\begin{tabular}{@{}llp{3cm}lp{3cm}@{}}
\toprule
Year & First Author      & Market Type                          & Application                   & Algorithm Used         \\ \midrule
2021 & Gomes I.L.R. \cite{Gomes} & Microgrid                            & Microgrid management          & K-Means Clustering     \\
2020 & Imran K.  \cite{Imran}   & International/National & Bilateral trading             & Bayesian classifier    \\
2017 & Čaušević S. \cite{Causevic} & International/National & Load scheduling               & Novel clustering algorithm             \\ \bottomrule
\end{tabular}
\caption{Articles relating to unsupervised learning algorithm type.}
\label{table:unsupervised-learning}
\end{table*}

\subsection{Optimisation}

In this subsection we review the papers which use optimisation as a basis for investigation.

Bevilacqua \textit{et al.}\cite{Bevilacqua} compare three optimisation methods to implement agent rationality in the Italian electricity market with an ABM. Through this, they observe that the moddel exhibits a very good fit to real data.

Duan \textit{et al.}\cite{Duan} propose a bi-level coordination optimisation integrated resource strategy to unify supply-side and demand side resources across China. They do this for mid-long term planning.

Gao \textit{et al.}\cite{Gao} use a genetic algorithm to determine an optimal bidding strategy.  They verify their approach by modelling a 30-bus system as an example. Meng \textit{et al.}\cite{Meng} also use a genetic algorithm optimisation approach to make dynamic pricing decisions. They model the day-ahead market and propose a two-level optimisation model. They model the price responsiveness of different customers using the optimisation algorithm, and then optimise the dynamic prices that the retailer sets to maximise its profit. They confirm the feasibility and effectiveness of their technique through simulation.

\begin{table*}[]
\caption{Articles relating to optimisation algorithm type.}
\begin{tabular}{@{}llp{3cm}lp{3cm}@{}}
\toprule
Year & First Author       & Market Type                          & Application        & Algorithm Used                                \\ \midrule
2019 & Bevilacqua S. \cite{Bevilacqua} & International/National & Agent behaviour    & Genetic Algorithm,Particle Swarm Optimisation \\
2018 & Duan W.  \cite{Duan}     & International/National & Expansion planning & Bi-level coordination optimization            \\
2018 & Gao Y. \cite{Gao}       & International/National & Bidding strategies & Genetic Algorithm                             \\
2018 & Meng F.   \cite{Meng}    & Local energy market                  & Tariff design      & Genetic Algorithm                             \\ \bottomrule
\end{tabular}
\label{table:optimisation}
\end{table*}

\subsection{Game theory}

Filho \textit{et al.}\cite{Filho} is the only paper in this review which takes a game theoretic approach. They deal with a comparative analysis of individual strategies of generating units in auctions, using non-cooperative game theory approach. They find that their method is best suited for second-price auctions and can be extended to more complicated networks with high precision. However, they find that it is not possible to use this methodology in some, more complex, systems. For example, bidding within a stochastic and uncertain environment, where a predictions are required for the behaviour of other actors.

\begin{table*}[]
\caption{Articles relating to game theoretic algorithm type.}
\begin{tabular}{@{}llp{3cm}lp{3cm}@{}}
\toprule
Year & First Author      & Market Type                          & Application        & Algorithm Used \\ \midrule
2016 & Filho N.S.C. \cite{Filho} & International/National & Bidding strategies & Game theory    \\ \bottomrule
\end{tabular}
\label{table:game-theory}
\end{table*}

\section{Conclusion}
\label{sec:future}

This paper provides a systematic review of artificial intellgigence (AI) applied to agent-based models within the electricity sector. Through this analysis, various research gaps have been identified. This is because, the majority of papers focus on similar themes and algorithms. This, therefore, highlights a wide range of areas in which the research community can focus on in future work.

Significant and increasing research interest  has been placed into agent-based models applied to the electricity sector. However, this work has shown that a lot of research has clustered around similar subjects. For instance, the top application investigated for each market type is bidding strategies, largely exploring whether it is possible to use reinforcement learning to bid into an electricity market. However, the impact of these bidding strategies on the wider market has been investigated to a lesser extent. Through classification of the different paper types, a long-tail was identified in research undertaken into different topics. For instance, a small amount of papers carried out work on the forecasting of carbon emissions, electricity grid control or bilateral trading. This, therefore, suggests that further work could be carried out to build upon this work in the future. 

Another aspect that has been explored to a lesser extent is the optimisation of the electricity system parameters as a whole as applied to policy. Much of the work is focused on micro-behaviours within the energy system. For example, improving on the demand response application. Kell \textit{et al.} optimise the whole energy system by modifying selected parameters\cite{Kell2020b} showing that this work can be carried out to great effect.

It was found that only 18\% of papers focus on the local electricity market. A significantly smaller proportion when compared to the international/national electricity market. This may be due to the lack of publicly available data for these local electricity markets. There therefore exists a high impact opportunity to collect and make local electricity market data available. This is because local electricity markets will become increasingly important for future systems with more small-scale renewable energy projects.

However, the wide-range of applications that were investigated shows the large impact that artificial intelligence can have on the electricity sector when combined with agent-based models. Due to the generalisable nature of these techniques findings from one application can be utilised to inform other applications. For instance, the highly researched bidding strategies domain can be used to inform microgrid management, which was only investigated by a single paper. Through this approach, it is suggested that high efficiencies can be gained within the power sector.

\section*{Acknowledgements}

This work was supported by the Engineering and Physical Sciences Research Council, Centre for Doctoral Training in Cloud Computing for Big Data [grant number EP/L015358/1].

\bibliographystyle{ACM-Reference-Format}
\bibliography{references/library,references/custombibtex-carbon-optimiser,references/custombibtex-elecsim-1,references/custombibtex-elecsim-poster,references/custombibtex-validation-optimiser,references/forecasting,references/forecasting-full,references/forecasting-note,references/custombibtexfile,references/custom_bib,references/bib_custom,references/custombibtex,references/ftt-power-custom,references/systematic-review}

\appendix

\end{document}